\documentclass{article}

\usepackage{emulateapj,epsfig,onecolfloat}

\newcommand{\msun}{\mbox{ M$_{\odot}$}}
\newcommand{\bq}{\begin{equation}}
\newcommand{\eq}{\end{equation}}
\newcommand{\kpc}{\mbox{ kpc}}
\newcommand{\pc}{\mbox{ pc}}
\newcommand{\mpc}{\mbox{ Mpc}}
\newcommand{\yr}{\mbox{ yr}}
\newcommand{\kel}{\mbox{ K}}
\newcommand{\cmden}{\mbox{ cm$^{-3}$}}
\newcommand{\gpcden}{\mbox{ Gpc$^{-3}$}}
\newcommand{\mpcden}{\mbox{ Mpc$^{-3}$}}

\newcommand{\yrinv}{\mbox{ yr$^{-1}$}}

\newcommand{\flux}{\mbox{ cm$^{-2}$  s$^{-1}$}} 
\newcommand{\fluxen}{\mbox{ cm$^{-2}$  s$^{-1}$ keV$^{-1}$}} 
\newcommand{\erg}{\mbox{ erg}}
\newcommand{\kms}{\mbox{ km s$^{-1}$}}
\newcommand{\keV}{\mbox{ keV}}

\newcommand{\amin}{\mbox{ arcmin}}

\begin{document}
 
\twocolumn
[ 
\title{Identifying Gamma-Ray Burst Remnants Through \\ Positron
Annihilation Radiation} 

\author{Steven R. Furlanetto \& Abraham Loeb}
\affil{Harvard-Smithsonian Center for Astrophysics, 60 Garden St.,
Cambridge, MA 02138;\\
sfurlanetto@cfa.harvard.edu, aloeb@cfa.harvard.edu}

\begin{abstract}
We model the annihilation of relic positrons produced in a gamma-ray burst
(GRB) after its afterglow has faded.  We find that the annihilation signal
from at least one GRB remnant in the Milky Way galaxy should be observable
with future space missions such as INTEGRAL and EXIST, provided that the
gas surrounding the GRB source has the typical density of the interstellar
medium, $\la 1 \cmden$.  Three fortunate circumstances conspire to make the
signal observable.  First, unlike positrons in a standard supernova, the
GRB positrons initially travel at a relativistic speed and remain ahead of
any non-relativistic ejecta until the ejecta become rarefied and the
annihilation time becomes long.  Second, the GRB remnant remains
sufficiently hot ($T \ga 5 \times 10^5 \kel$) for a strong
annihilation line to form without significant smearing by three-photon
decay of positronium.  Third, the annihilation signal persists over a
time longer than the average period between GRB events in the Milky
Way galaxy. 
\end{abstract}

]
\section{ Introduction }

The detailed nature of the central engine of gamma-ray bursts (GRBs) is
still unknown (van Paradijs, Kouveliotou, \& Wijers 2000; Piran 2000,
and references therein). Attempts to derive empirical constraints on
the environment (Djorgovski et al. 2001, and references therein), the
frequency (Schmidt 2001), the collimation angles and energy output
(Frail et al. 2001; Freedman \& Waxman 2001; Panaitescu \& Kumar
2001b; Piran et al. 2001) or the possible association of GRBs with
supernovae (Bloom et al. 1999; Kulkarni et al. 2000; Reichart 
2001) are compromised by the difficulty of observing GRBs across large
cosmological distances. Obviously, identification of old GRB remnants in
the local universe would provide much better insight into the nature of GRB
progenitors (Loeb \& Perna 1998; Woods \& Loeb 1999; Perna, Raymond \& Loeb
2000; Paczynski 2001; Ayal \& Piran 2001).

Given the recently inferred similarity between the energy output in
supernovae and GRBs, $\sim 10^{51}~{\rm ergs}$ (Frail et al. 2001), it now
appears difficult to separate the late evolution stages of their remnants
hydrodynamically.  Even if a GRB explosion is initially highly beamed, the
asymmetry of the blast wave it generates in the surrounding interstellar
medium (ISM) would be erased on an isotropization timescale $t_{\rm iso}
\sim 6 \times 10^3 (E_{51}/n_0)^{1/3} \yr$ (Ayal \& Piran 2001), 
where $E_{51}$ is the kinetic energy output of the GRB in units of $10^{51}
\erg$ and $n_0 = \rho_{\rm ISM}/\mu_b m_p$ is the ambient number density of
atoms in units of ${\rm cm}^{-3}$ for a mean molecular weight $\mu_b =
1.4$.  Subsequently, the blast wave expands just as in a normal
supernova (SN) remnant. Any non-relativistic, SN-like ejecta would
moderate the initial GRB asymmetry as soon as it overtakes the
decelerating GRB shock (Piran \& Ayal 2002). External anisotropy may also
result from the interaction of the remnant blast wave with a
non-uniform ISM.  Thus, morphological studies alone cannot
unambiguously identify GRB remnants.  The alternative method of
seeking spectral signatures of the photoionized regions around GRBs
(Perna, Raymond, \& Loeb 2000) also suffers from potential confusion
with SNe occurring in unusual environments.

In this {\it Letter}, we point out that positrons produced during the early
relativistic GRB phase offer a powerful tool for identifying GRB remnants.
We model the expansion of a GRB remnant and show that a well-defined
annihilation line should be visible until radiative cooling becomes
important, at which time the remaining positrons annihilate rapidly (see
also Dermer \& B{\" o}ttcher 2000).  

\section{Model of GRB Remnants}

We model the initial GRB explosion as two highly-relativistic
jets of material moving in opposite directions, initially covering a
fraction $f_b$ of a sphere surrounding the GRB.  We assume that a fraction
$\xi_+$ of the GRB energy is transformed into $e^+$ \emph{rest mass}.  The
positrons can either be produced during the burst itself (Cavallo \&
Rees 1978; Shemi \& Piran 1990) or by interactions of the
$\gamma$-rays with the ambient medium (Thompson \& Madau 2000; Dermer
\& B{\" o}ttcher 2000; M{\' e}sz{\' a}ros, Ramirez-Ruiz, \& Rees
2001).  We scale our results to a fiducial value of $\xi_{-2.3} =
(\xi_+/0.005)$ corresponding to half of the GRB energy in positrons
with a bulk Lorentz factor $\gamma \sim 100$ as required by the
compactness argument (Piran 2000).  We therefore assume a total of
$N_+ = 6 \times 10^{54} \xi_{-2.3} E_{51}$ positrons in the GRB. 

The positrons initially travel at a relativistic speed just ahead of the
GRB jet.  The jet begins to decelerate when it reaches a radius $R_j \sim
0.2 (E_{51}/n_0)^{1/3} \pc$ (Rhoads 1997).  At this point, it begins to
expand sideways and slows exponentially to non-relativistic speeds.
Thereafter, simulations have shown that $R_j \propto t^{1/3}$ until the
remnant isotropizes at $t_{\rm iso}$ and begins a Sedov-Taylor spherical
expansion (Ayal \& Piran 2001).

If the commonly hypothesized link between GRBs and SNe exists, then the jet
will be followed by SN ejecta of mass $M_{\rm ej} \sim 10 \msun$.  The
ejecta expand freely behind the jet and eventually impact the
(decelerating) positrons at a time $t_{\rm mix} \sim 110
(E_{51}/n_0)^{1/3} v_4^{-3/2} \yr$, where $v_4$ is the ejecta velocity
in units of $10^4 \kms$ (Piran \& Ayal 2002).  At this point, the density
is $n_{\rm mix} \ga 50 M_1 (n_0/E_{51}) v_4^{3/2} \cmden$, assuming
that the ejecta are distributed uniformly throughout the remnant and where
$M_1=(M_{\rm ej}/10 \msun)$.  If mixing of the ejecta and shocked gas is
efficient, then the ejecta-dominated expansion models of Truelove \&
McKee (1999) show that by this point the thermal energy imparted to
the swept-up material could heat the remnant to a temperature $T_{\rm
mix} \ga 10^8 \kel~ (E_{51}/M_1)$.  The ratio of the annihilation time to
remnant age does not change with time, $\tau_{\rm ann}/t \sim 500
M_1^{-1} (E_{51}/n_0) v_4^{-3/2}$.  Thus, only a small fraction of the
positrons should annihilate during this very early stage, even if the
ejecta are concentrated in a shell of width $\sim 0.1 R_{\rm mix}$.

This contrasts with the survival of positrons in a standard SN, where a
substantial number of positrons can be created through the decay chain
$^{56}$Ni$\rightarrow ^{56}$Co$\rightarrow ^{56}$Fe$ + e^+$, which has a
decay time of $\tau \sim 111$ days and occurs for $18\%$ of the $^{56}$Ni
atoms.  [In SN1987A, observations imply that $\sim 2.6 \times 10^{53}$
$e^+$ were produced through this channel (Suntzeff \& Bouchet 1990).]  The SN
positrons are produced inside the cool, dense SN ejecta when $\tau_{\rm
ann}/t \la 0.03$ (depending on the uncertain ionization and temperature
structure of the ejecta).  Therefore, most of the SN positrons are
annihilated within a few years of the SN explosion.  Milne, The, \&
Leising (1999) have confirmed numerically that only a small fraction
of the positrons can escape the initial stages of the remnant
evolution in Type Ia SNe. 

After passing through the external shock most positrons will rapidly
thermalize with the ambient medium while a small fraction ($\sim 10^{-3}$)
will be re-accelerated by the (collisionless) external shock
(Gieseler, Jones, \& Kang 2000).  If positrons are produced ahead of
the shock, then they will have a characteristic Lorentz factor
$\gamma_+ \sim 30$ (M{\' e}sz{\' a}ros et al. 2001).  Regardless of
their initial state, the positrons cool rapidly through Coulomb
collisions and adiabatic decompression as the remnant expands.  While
they are relativistic, adiabatic cooling leads to energy loss on the
dynamical timescale of the remnant, $\tau_{\rm ad} \sim R_s/v_s $,
where $R_s$ and $v_s$ are the radius and expansion speed of the 
(assumed spherical) remnant.  Once the positrons become non-relativistic,
their Coulomb cooling time is short, $\tau_{\rm coul} \sim 3 \times 10^4
(v_+/c) n_0^{-1} \yr$, where $v_+$ is the positron velocity.  Using the
analytic expressions for the cooling rates in Furlanetto \& Loeb
(2002) supplemented by expansion cooling, we have confirmed that the
cooling timescales for $\gamma_+ \la 100$ are much shorter than the
duration of the Sedov phase.  We therefore simply assume that all
positrons begin in thermal equilibrium with the shocked ambient gas.

Once the remnant isotropizes, it follows the Sedov (1959) solution until
radiative cooling becomes important.  Although the remnant is initially
asymmetric because of the jet geometry, Ayal \& Piran (2001) have
shown numerically that even before isotropization, the total volume of
the remnant scales with time similarly to a Sedov self-similar blast
wave.  For this reason, and because $t_{\rm iso}$ is much smaller than
the duration of the adiabatic phase, we treat the remnant boundary as
a Sedov blast wave throughout both of these phases.  Then $R_s = 0.314
(E_{51}/n_0)^{1/5} t_{\rm yr}^{2/5} \pc$, where $t_{\rm yr}$ is the
remnant age in years.  The adiabatic index of the ambient medium is
taken to be $\Gamma = 5/3$. 

To calculate $N_+(t)$, we assume that that the positrons are perfectly
mixed throughout the interior of the remnant, so that the $e^+$ density
$n_+(r) \propto n_e(r)$, where $n_e$ is the free electron density.  We
expect such mixing to occur on a timescale $\sim t_{\rm iso}$ or faster
because of instabilities on the interface between the SN ejecta and the
ambient medium (Chevalier, Blondin, \& Emmering(1992); Jun, Jones, \&
Norman 1996).  The annihilation rate is then 
\bq 
\dot{N}_+ = - \int_0^{R_s} n_e(r) \, n_+(r) \alpha[T(r)] 4 \pi r^2 \, dr
\label{eq:Ndot}
\eq 
where the radial distributions of $n_e$ and $T$ are given by
Sedov (1959).  Here $\alpha = \alpha_{\rm fa} + \alpha_{\rm Ps}$ is the
annihilation rate coefficient with contributions from two channels.  The
first component, $\alpha_{\rm fa}$, involves direct annihilation with free
electrons.  In the non-relativistic regime ($T \la 10^8 \kel$), we use the
rate coefficient given by Gould (1989) which includes corrections due to
Coulomb focusing.  Above this temperature we use the fitting formula given
by Svensson (1982).  The second coefficient $\alpha_{\rm Ps}$ describes
radiative recombination to positronium (Ps) and is given by Gould (1989).
Although the corresponding expression is strictly valid only in the
non-relativistic regime, Ps formation is sub-dominant for $T_{\rm ann} >
10^6 \kel$ and so the errors introduced by applying it to high temperatures
are negligible.  The Ps lifetime is $\tau_{\rm Ps} < 1.33 \times 10^{-7}
\sec$ (depending on the energy level to which recombination occurs), and so
we assume that positrons annihilate immediately after forming Ps.

Once the shocked gas cools radiatively to $T \la 10^4 \kel$, a dense shell
begins to form at the shock front. We use the cooling function of
Cioffi, McKee, \& Bertschinger (1988), for which the nominal shell
formation time is $t_{\rm sf} = 3.61 \times 10^4 E_{51}^{3/14}
n_0^{-4/7} \yr$.  The shell formation process is not fully understood,
but two-dimensional simulations have shown it to be both unstable and
violent (Blondin et al. 1998).  These simulations also 
find that the dense shell does not fully form until $\sim 1.5 t_{\rm sf}$.
We therefore show results up to this time with the caveat that once cooling
becomes important at $t_{\rm sf}$ they may no longer be accurate.  After
shell formation we expect the positrons to annihilate rapidly.  Not only
does the density of the shell increase dramatically, but also charge
exchange with HI comes to dominate as the temperature falls below $\sim
10^5 \kel$ (Bussard, Ramaty, \& Drachman 1979; see \S \ref{disc}).
Thus, we do not expect the annihilation signal to persist after the
Sedov phase ends. 

Even before shell formation, the loss of energy to cooling radiation
decelerates the remnant.  We use the analytic approximation of
Cioffi et al. (1988) to describe the blast wave size and velocity in
this phase (for $t > t_{\rm sf}/e$, where $e$ is Euler's constant).
As there is no analytic understanding of the interior structure of the
remnant at this stage, we continue to use profiles from the Sedov
(1959) solution, normalized to the total size and velocity of Cioffi
et al. (1988).  Our results are only weakly dependent on the details
of these solutions. 

\section{ Results }

Figure \ref{fig:flux} shows the expected flux of annihilation photons,
$F_\gamma$, as a function of time for blast waves traveling through media
with densities $n_0 = 10^3, 10^2, 10^1, 1$, and $0.1 \cmden$ (from top to
bottom).  For each density we carry the integration through $1.5 t_{\rm
sf}$.  The solid lines show the photon flux from direct annihilation
while the dashed lines show that from Ps formation.  We scale our results
to $d_{10}$, the remnant distance in units of $10 \kpc$.  The photon
production rate is nearly proportional to the ambient density because at a
fixed temperature $\tau_{\rm ann} \propto n_e^{-1}$.  We find that the
average temperature of annihilating positrons is $T_{\rm ann} \approx 1.25
T_s$, where $T_s$ is the postshock temperature.  The annihilation weighted
density is $n_{\rm ann} \approx 2.7 n_0$.  Both of these formulae are valid
to within $\sim 5\%$ for all times and densities.  Thus our model finds
that most of the annihilations occur in a shell near the shock front.
Throughout most of the remnant lifetime, $5 \times 10^5 \kel \la
T_{\rm ann} \la 10^7 \kel$.  Thus, direct $e^+e^-$ annihilation
dominates, with Ps formation becoming important only at late times
when the shock velocity has fallen substantially.  Note that
$F_{\gamma}$ is approximately proportional to $E_{51}$, but the
explosion energy also affects $t_{\rm sf}$. 

\begin{center}
\epsfig{file=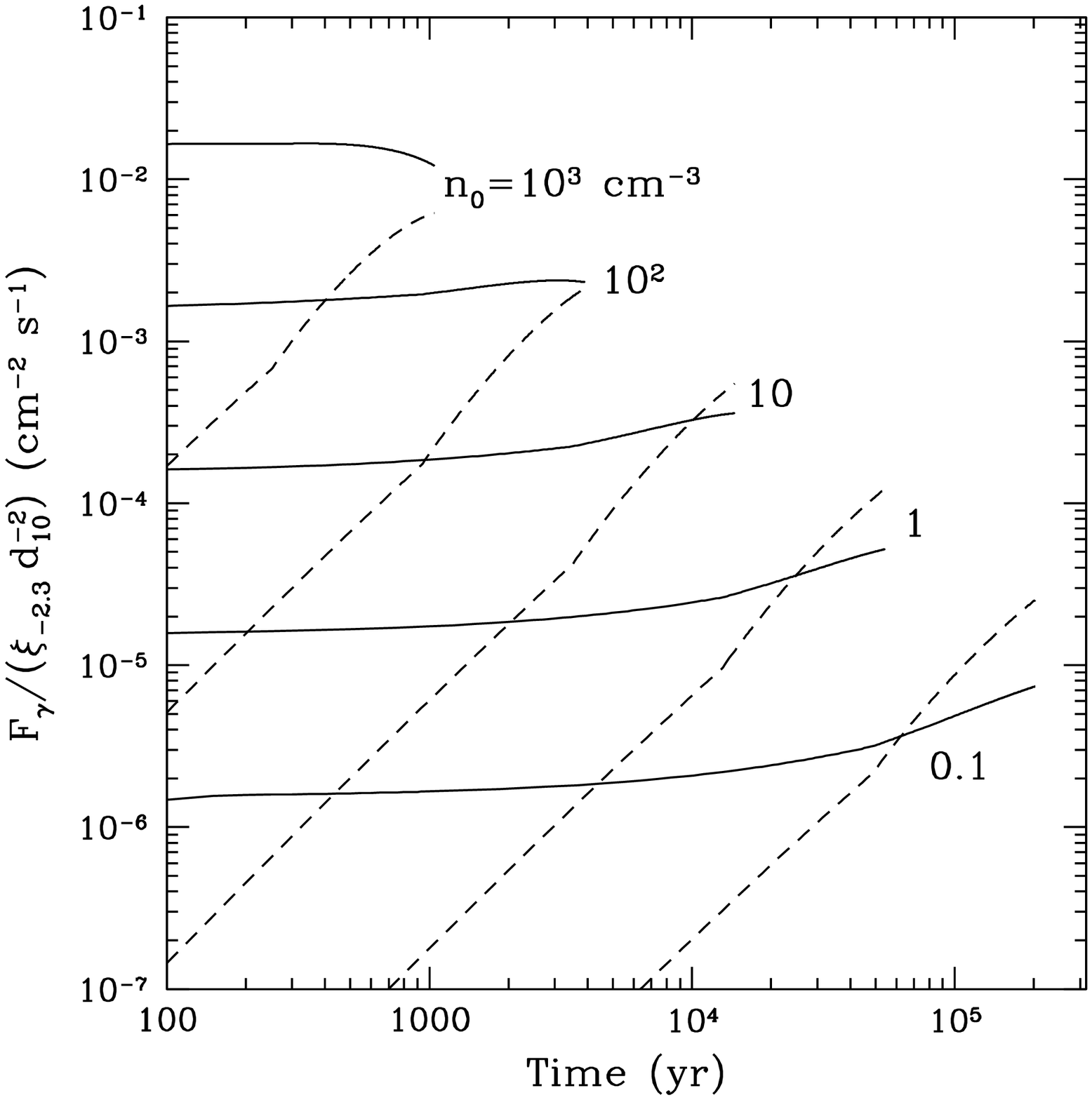,width=8.4cm,height=7.56cm}
\figcaption{ Flux of annihilation photons as a function of time for remnants
embedded in a medium of density $n_0 = 10^3,10^2,10,1$, and $0.1 \cmden$,
from top to bottom.  Solid curves show the photon flux from free
annihilation, while dashed curves show the flux from Ps decay.  All results
are scaled to a remnant distance $d_{10}=(d/10 \kpc)$ and assume
$E_{51}=1$.  
\label{fig:flux} }
\end{center}

We next calculate the emission spectrum of the remnant.  The spectrum
produced by direct annihilation when $k_B T_e \ll m_e c^2$ is given by
Svensson, Larsson, \& Poutanen (1996).  We ignore Coulomb corrections
to the annihilation rate, which is a good assumption for the
temperatures of interest here (Gould 1989).  The decay of Ps is more
complicated.  One-quarter of the recombinations produce para-Ps, which
annihilates into two photons with the spectral distribution given by
Gould (1989).  Three-quarters of the recombinations produce ortho-Ps,
which must decay into three photons in order to conserve angular
momentum.  The spectrum in this case is a continuum, with a form in
the Ps rest frame given by Ore \& Powell (1949).  We 
include the broadening due to the thermal motions of the plasma particles
by transforming to the observer frame (Bussard et al. 1979) and using the
relativistic Maxwell-Boltzmann velocity distribution.\footnote{Note,
however, that our expression for $\alpha_{\rm Ps}$ does not include
relativistic corrections, so our continuum level may be an overestimate at
very early times when $T_{\rm ann} \gg 10^8 \kel$.} In calculating the
spectra, we assume for simplicity that all annihilations take place at a
single temperature $T_{\rm ann}$.  We also neglect the broadening
introduced by the expansion speed of the remnant, because the ratio between
the thermal $e^+$ speed and the shock speed is $v_{\rm th}/v_s \sim (m_p/2
m_e)^{1/2} \gg 1$, where $m_e$ and $m_p$ are the electron and proton
masses.

Snapshots of the annihilation spectrum at $t/t_{\rm sf} = 0.3, 0.6, 0.9,
1.2$, and $1.5$ are shown in Figure \ref{fig:spec}, for media with $n_0 = 1
\cmden$ (top panel) and $n_0 = 10^3 \cmden$ (bottom panel).  As shown in
Figure \ref{fig:flux}, the number of free annihilations per second varies
relatively slowly with remnant age, increasing primarily because of the
decreasing $T_{\rm ann}$.  The line also narrows with time: the full width
at half maximum is $\sim 30 (T_{\rm ann}/10^7 \kel)^{1/2} \keV$.  The line
is narrower for lower density media at constant $t/t_{\rm sf}$ because
$v_s$ and hence $T_{\rm ann}$ are smaller.

\begin{center}
\epsfig{file=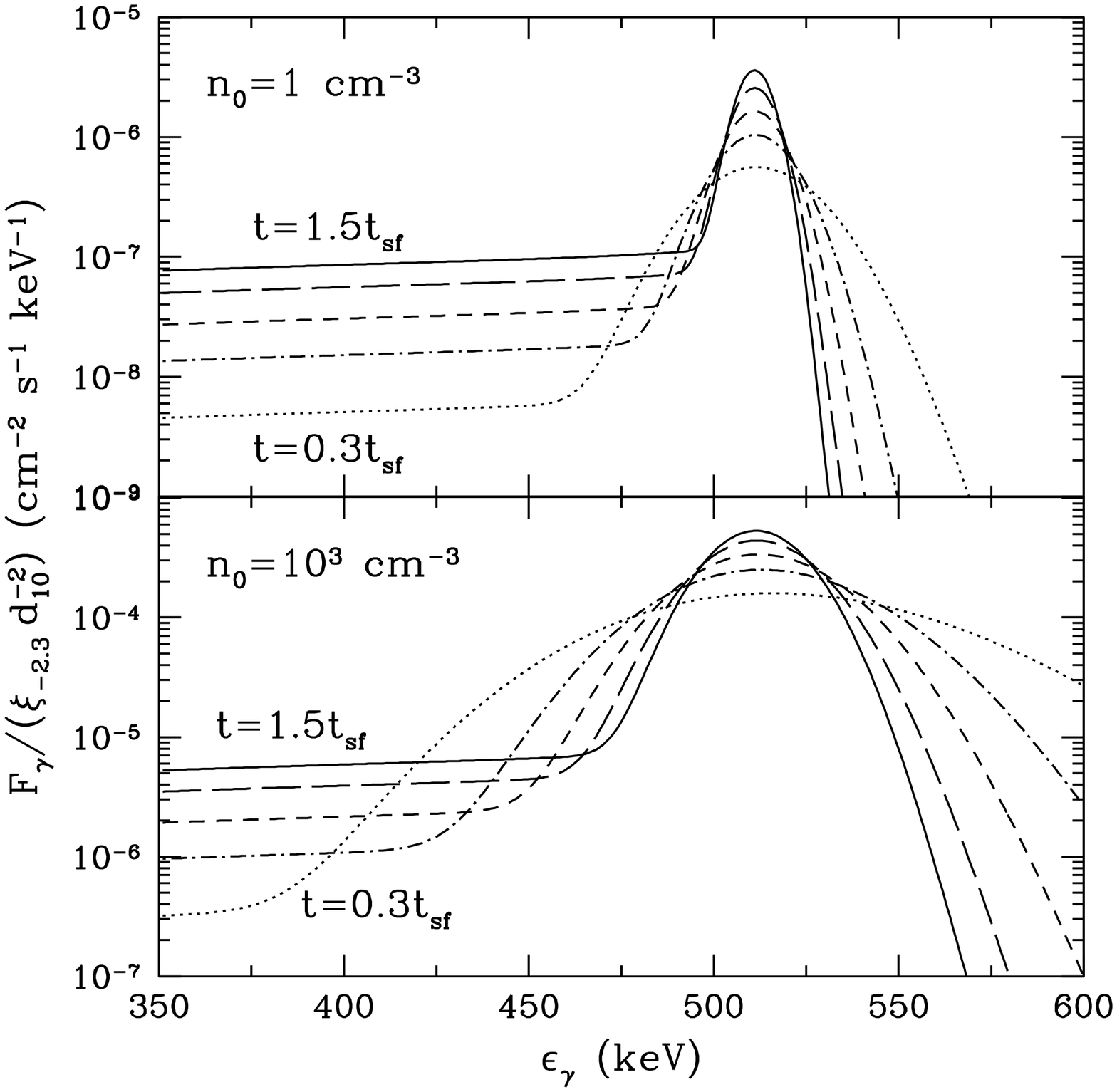,width=8.4cm,height=7.56cm}
\figcaption{ Spectral fluxes as a function of photon energy at various times.
Results are shown for $n_0=1 \cmden$ (\emph{top panel}) and $n_0=10^3
\cmden$ (\emph{bottom panel}) at times $t/t_{\rm sf}=0.3$ (\emph{dotted
curves}), $0.6$ (\emph{dot-dashed curves}), $0.9$ (\emph{short-dashed
curves}), $1.2$ (\emph{long-dashed curves}), and $1.5$ (\emph{solid
curves}).  All results are scaled to a remnant distance $d_{10}=(d/10
\kpc)$ and assume $E_{51}=1$.  The asymmetric line broadening is a property
of the annihilation spectrum given by Svensson et al. (1996). 
\label{fig:spec} }
\end{center}

\section{ Discussion }
\label{disc}

The BATSE data set implies a GRB rate per comoving volume of $\dot{n}_{\rm
iso} \sim 0.5 \gpcden \yrinv$, assuming isotropic emission (Schmidt 2001).
The actual event rate is estimated to be larger by an average beaming
factor, $\langle f_b^{-1} \rangle \sim 520$ (Frail et al. 2001).  Most models
associate GRBs with compact stellar remnants; under the assumption that the
GRB event rate is proportional to the star formation rate we can convert
the above rate density to an event rate per unit mass of star formation
(e.g., Porciani \& Madau 2001).  The expected period between GRB
events in the Milky Way is then $\tau_{\rm MW} \sim 3 \times 10^4 \yr~
(\dot{n}_{\rm iso}/0.5)^{-1} (\langle f_b^{-1} \rangle/520)^{-1}
(R_0/0.007) R_{\rm MW}^{-1}$, where $R_0$ and $R_{\rm MW}$ are the
present day average star formation rates in the universe and in the
Milky Way in units of ${\rm M}_{\odot} \mpcden \yrinv$ and ${\rm
M}_{\odot} \yrinv$, respectively, and $\dot{n}_{\rm iso}$ has units of
${\rm Gpc}^{-3} \yrinv$. 

Together with Figure \ref{fig:flux}, this suggests that the positron
annihilation line should be detectable from at least one Galactic GRB
remnant at any time provided that the circumburst medium has $n_0 \la 1
\cmden$.  This conclusion relies on three fortunate coincidences.  First
and most importantly, the shell formation time satisfies coincidentally
$t_{\rm sf} \sim \tau_{\rm MW}$ for $n_0 = 1 \cmden$.  Although some
fraction of GRBs may occur in dense molecular clouds, recent modeling of
jetted afterglows indicates that the ambient density is in the range $n_0
\sim 10^{-3}$--$10^{1.5} \cmden$ (Berger et al. 2000, 2001; Harrison
et al. 2001; Panaitescu \& Kumar 2001a).  Second, in contrast to $e^+$
created by SNe, GRB positrons annihilate on a timescale $\tau_{\rm
ann} \ga t_{\rm sf}$.  Third, $T_{\rm ann} \ga 5 \times 10^5 \kel$, so
that a clear line from the direct annihilation channel is visible
throughout the source lifetime. 

Two backgrounds could contaminate the annihilation line signal.  First,
relativistic protons accelerated by the collisionless shock can create
positrons through the decay of $\pi^+$ produced in collisons with thermal
protons.  Assuming that a fraction $\xi_p$ of the GRB energy is invested in
the acceleration of protons (Blandford \& Eichler 1987) and using the
approximate interaction cross-sections of Mannheim \& Schlickeiser
(1994), we find that the number of positrons produced before shell
formation is $N_{+,\,\pi} \sim 10^{48} E_{51}^{17/14} n_0^{3/7}
(\xi_p/0.1)$ (see Furlanetto \& Loeb 2002).  Unless $\xi_{-2.3} \la
10^{-6}$, the positron population from this process is negligible
compared to that produced in the GRB.  A second background is due to
inverse-Compton (IC) emission by the electrons accelerated at the
shock.  The recent detection of TeV $\gamma$-ray emission from SN
remnants (Tanimori et al. 1998; Muraishi et al. 2000) coupled with the
well-observed high-energy synchrotron emission from the same remnants
(Koyama et al. 1995, 1997) provides strong evidence for the existence
of this population of electrons and calibrates the IC signal.  If we
extrapolate the observed TeV $\gamma$-ray flux to the energy of the
annihilation line using the spectral index of the radio synchrotron
emission (which should also be the index of the IC spectrum; Reynolds
2001), we find a flux $F_{\gamma,\,{\rm IC}} \la 10^{-12} d_{10}^{-2}
(\epsilon_\gamma/511 \keV)^{-1.5} \fluxen$.  Thus, the IC background
should be negligible compared to the annihilation signal. 

The prospects for detection of young GRB remnants in the Milky Way galaxy
are good, provided that GRB sources are not confined to dense molecular
clouds.  As shown in Figure \ref{fig:flux}, a GRB remnant produces a
characteristic annihilation line photon flux of $F_\gamma \sim 3 \times
10^{-5} \xi_{-2.3} n_0/d_{10}^2 \flux$.  The INTEGRAL
satellite,\footnote{See
http://astro.estec.esa.nl/SA-general/Projects/Integral/integral.html}
expected to be launched in October 2002, will have spectral capabilities in
the energy range of interest.  For a $10^6 \sec$ observation, its SPI and
IBIS instruments have $3 \sigma$ line sensitivities of $\sim 5 \times
10^{-6} \flux$ and $\sim 2 \times 10^{-5} \flux$, respectively.
EXIST,\footnote{ See http://exist.gsfc.nasa.gov } a proposed all-sky hard
X-ray survey mission, has an expected $5 \sigma$ line sensitivity of $\sim
5 \times 10^{-6}\flux$ in the relevant energy range (assuming an
integration time of $10^7 \sec$, the mean exposure time planned for any
point on the sky).  These instruments will also have the spectral
resolution to identify the line.  The IBIS instrument on INTEGRAL and the
EXIST survey, with angular resolutions of $12\arcmin$ and $5\arcmin$
respectively, will also be able to map nearby GRB remnants, for which the
angular size at $t_{\rm sf}$ is $\sim 14 E_{51}^{2/7} n_0^{-3/7}
d_{10}^{-1} \amin$.

If GRBs occur primarily in dense environments, $t_{\rm sf}$ decreases and
the probability for observing a remnant declines.  Figure \ref{fig:flux}
shows that the average flux in the annihilation line from a nearby
extragalactic source ($d \ga 1 \mpc$) during the Sedov phase will
likely remain below the sensitivity limits of the above instruments.
However, once shell formation begins, the positrons will annihilate through
charge exchange with recombining hydrogen on a timescale $\tau_{\rm cx}
\sim 10/(f_{\rm HI} n_{\rm sh}) \yr$, where $f_{\rm HI}$ is the neutral
fraction (Bussard et al. 1979).  If the shell is supported primarily
by magnetic and cosmic ray pressure, Cox et al. (1999) find that the
shell density at $t_{\rm sf}$ is $n_{\rm sh} \sim 100 B_0^{-1}
n_0^{23/14} E_{51}^{1/14} \cmden$, where $B_0$ is the transverse
component of the ambient magnetic field in units of $\mu {\rm G}$.
The maximal annihilation flux is then $F_{\gamma,{\rm sh}} \sim 3
\times 10^{-3} f_{\rm HI} \xi_{-2.3} E_{51}^{15/14} n_0^{23/14}
B_0^{-1} d_{\rm Mpc}^{-2} \flux$, where $d_{\rm Mpc}$ is the distance
to the GRB in Mpc.  Because the charge exchange process produces Ps,
$75\%$ of the annihilation energy is released in the continuum, but
this suggests that the remnants will brighten significantly during the
brief phase of shell formation and may be visible out to the Virgo
cluster distance, $d \sim 20 \mpc$.  Given the number of galaxies in
Virgo, we would therefore expect to find at least one Virgo remnant in
this stage provided that $f_{\rm HI} n_{\rm sh} \la 1 
\cmden$.  The details of the luminosity and spectrum will depend critically
on modeling the process of shell formation and cooling.  Such modeling
stretches the limits of current simulations (e.g., Blondin et
al. 1998; Cox et al. 1999) but may become feasible in the future.

\acknowledgements

We thank J. Raymond for helpful discussions.  This work was supported in
part by NASA grants NAG 5-7039, 5-7768, and NSF grants AST-9900877,
AST-0071019 for AL.  SRF acknowledges the support of an NSF graduate
fellowship.

\end{document}